\documentclass[aps,pra,reprint,superscriptaddress,floatfix]{revtex4-2}

\usepackage{amsmath}
\usepackage{amssymb}
\usepackage{amsfonts}
\usepackage{graphicx}
\usepackage{dcolumn}
\usepackage{bm}
\usepackage{physics}

\begin{document}

\title{Saturation of the Cram\'er-Rao Bound for the Atomic Resonance Frequency with Phased Array of Hyperbolic Secant Pulses}

\author{Tharon Holdsworth}
\email[]{tharonholdsworth@tamu.edu}
\affiliation{Department of Physics and Astronomy, Texas A\&M University, College Station, Texas 77843, USA}\affiliation{Institute for Quantum Science and Engineering, Texas A\&M University, College Station, Texas 77843, USA}

\author{Jacob Adamczyk}
\affiliation{Department of Physics, University of Massachusetts Boston}

\author{Girish S. Agarwal}
\affiliation{Department of Physics and Astronomy, Texas A\&M University, College Station, Texas 77843, USA}\affiliation{Institute for Quantum Science and Engineering, Texas A\&M University, College Station, Texas 77843, USA}\affiliation{Department of Biological and Agricultural Engineering, Texas A\&M University, College Station, Texas 77843, USA}

\date{\today}

\begin{abstract}
Precise estimation of the atomic resonance frequency is fundamental for the characterization and control of quantum systems. The resonance experiment is a standard method for this measurement, wherein the drive field frequency is swept to invert the system population.  We analyze the classical and quantum Fisher information for the resonance experiment driven by hyperbolic secant shaped $\pi$-pulses; setting a fundamental limit on the precision obtainable using the resonance method. We show that measurements using sequences of pulses with alternating phases globally saturates the quantum Cram\'er-Rao bound, achieving the theoretical limit of precision for atomic resonance frequency estimation. 
\end{abstract}


\maketitle

\section{Introduction}

Quantum sensors exploit properties of quantum systems to measure physical parameters with precision exceeding that of classical measurement devices. The field of quantum sensing has developed to study the capability and deployment of quantum sensors~\cite{davidovich2024quantum}. A canonical task in quantum sensing is the measurement of the atomic resonance frequency between two discrete energy levels of a quantum system. The first method introduced for this purpose is the Rabi molecular beam resonance experiment. The Rabi method subjects a two state system to a frequency-swept field until the resonance condition is achieved, producing the maximum measured response~\cite{rabi1938new}. Ramsey improved on the Rabi method with the method of separated oscillatory fields that introduced a second free procession time alongside the interaction time \cite{ramsey1990experiments}. 
Such resonance experiments form the foundation of many quantum sensing applications, including the operation of atomic clocks \cite{ludlow2015optical} and nitrogen vacancy in diamond magnetometry~\cite{dreau2011avoiding,levine2019principles,barry2020sensitivity,degen2017quantum}.

A common metric of precision for the atomic resonance measurement is the spectral line width, however the linewidth is only a proxy for the precision of the measurement and depends on experimental details. A more rigorous analysis of the probability distributions, obtained from a discrete set of measurement outcomes, establishes the Fisher information as a lower bound on the variance of any unbiased estimator of the atomic resonance frequency. This bound, known as the Cramér-Rao bound, sets a limit on parameter estimation precision \cite{kay1993fundamentals}.  The framework of parameter estimation naturally extends to quantum systems, where the Fisher information forms a metric on the space of density operators, known as the quantum Fisher information \cite{braunstein1994statistical}. In this setting, the quantum Fisher information can be used to engineer arrangements where small variations in a parameter's value causes a large deviation in the measured response, thereby increasing measurement precision~\cite{ozawa2018extracting,pang2017optimal}. Consequently, quantum parameter estimation enables the design of measurement protocols that surpass classical precision limits utilizing non-classical resources \cite{giovannetti2006quantum, giovannetti2011advances,liu2020quantum}. 

With the fundamental lower bound on precision set, the identification of protocols that achieve or asymptotically approach the Cramér-Rao bound remains a central objective in quantum sensing \cite{yu2022quantum,paris2009quantum}. Saturation of this bound identifies an optimal measurement strategy under fixed resource constraints that inform the design of measurements schemes that maximize experimental sensitivity and efficiency.

In this letter, we analyze the atomic resonance experiment through the framework of quantum Fisher Information to determine conditions under which the Cram\'er-Rao bound is saturated.  Section~\ref{Section II} establishes notation and Fisher information analysis of the single pulse resonance measurement. Section~\ref{Section III} extends this analysis to a phase-modulated multi-pulse resonance measurement that is shown to saturate the Cram\'er-Rao bound in resonance frequency estimation.

\section{Single Pulse Resonance}\label{Section II}

Consider the Hamiltonian

\begin{align}
    H = H_0 - \Vec{\mu} \cdot \vec{E}(t),
\end{align}

where the two-level system Hamiltonian is ${H_0 = \frac{1}{2}\hbar \omega_0 \sigma_z}$, $\omega_0$ is the atomic resonance frequency, $\vec{\mu}$ is the dipole moment operator and electric field $\vec{E}(t) = \mathfrak{Re}(\hat{e}E_0(t)e^{i\omega t})$ \cite{allen2012optical}. In units of $\hbar = 1$, the Schr\"odinger equation for the above Hamiltonian within the rotating wave approximation reads

\begin{align}\label{Schrodinger Equation}
   i\partial_t\ket{\psi(t)} = -\frac{1}{2}\left(\Delta(t)\sigma_z  +\Omega(t)\sigma_x\right)\ket{\psi(t)}.
\end{align}

The Rabi frequency is defined to be $\Omega(t) = \abs{\hat{e}\cdot \vec{\mu}} E_0(t)$ and $\Delta(t)$ is the detuning function. We will study the hyperbolic secant pulse envelope defined by the functions

\begin{align}\label{Pulse Functions}
    \Omega(t) = \Omega_0 \sech(\frac{t}{\tau}),\\
    \Delta(t) = \Delta = \omega - \omega_0,
\end{align}

where $\Omega_0$ is the pulse amplitude, $\Delta$ is the static detuning, and $\tau$ is the characteristic time of the pulse.  This model is exactly solvable and the solutions have been studied by Rosen-Zener\cite{rosen1932double}, Allen-Eberly~\cite{allen2012optical}, McCall-Hahn~\cite{PhysRev.183.457} and Carrol-Hioe~\cite{hioe1984solution,hioe1985two}. With the initial condition $\ket{\psi( t = -\infty)} = \ket{0}$, the final state $\ket{\psi(t = \infty)}$ can be obtained from evolution under the unitary propagator  

\begin{align}\label{Single Pulse Propagator}
    U = \begin{pmatrix}
        a & -b^*\\
        b & a^*
    \end{pmatrix},
\end{align}

with the matrix elements 

\begin{subequations}\label{Cayley-Klein Parameters}
\begin{eqnarray}
a = \frac{\Gamma(\nu)\Gamma(\nu-\lambda-\mu)}{\Gamma(\nu-\lambda)\Gamma(\nu-\mu)},\label{a}\\
    b = \frac{\sqrt{-\lambda\mu}}{\abs{1-\nu}}\frac{\Gamma(2-\nu)\Gamma(\nu-\lambda-\mu)}{\Gamma(1-\lambda)\Gamma(1-\mu)},\label{b}
\end{eqnarray}
\end{subequations}

where $\Gamma$ denotes the Gamma function and the parameters $\lambda = \frac{\tau\Omega_0}{2}, \mu = -\frac{\tau\Omega_0}{2}, \nu = \frac{1}{2}(1 - i\tau \Delta)$.

\begin{figure}
    \centering
    \includegraphics[width=0.50\linewidth]{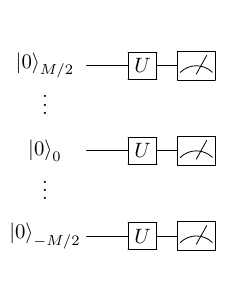}
    \caption{Quantum circuit diagram of the single pulse resonance measurement. Horizontal lines represent an experiment performed on a two-level probe system, squares represent interactions with the control field performed sequentially from left to right and a final projective measurement by the meter symbol. Experiments are performed by preparing the probe in the ground state, interacting with a single hyperbolic secant shaped $\pi$-pulse indexed by $\omega$ and a measurement of the population of the probe in the ground state. The index of the experiment with a final state closest to the ground state yields an estimate of the atomic resonance frequency $\omega_0$.}\label{fig:Single Pulse Sensing Circuit}
\end{figure}

When the resonance condition $\omega = \omega_0$, is satisfied with pulse area $\tau\Omega_0\ = 2n+1$ ($n \in \mathbb{Z}$),  the population is completely inverted from the ground state to the excited state. We will exploit this special case to estimate the value of the atomic resonance $\omega_0$. The measurement illustrated in figure~\ref{fig:Single Pulse Sensing Circuit} proceeds by preparing a qubit probe in the ground state and evolving the state under $U$ with $\Omega_0\tau = 1$ and $\omega$ in the interval $[\omega_{-M/2},\omega_{M/2}]$. The protocol is repeated for all $M$ values of $\omega$ by resetting the same probe system. Assuming $\omega_0$ is within the sampled interval, the population will completely invert near $\omega_{min} = \text{argmin}(\abs{a}^2)$, and the atomic resonance frequency is estimated to be $\omega_0  = \omega_{min} $.

The precision of the resonance method is quantified by the variance of $M$ independent repetitions of a single experiment. The variance of any unbiased estimator of the atomic resonance frequency $\hat{\omega}_0$, is lower bounded by the Cram\'er-Rao inequality~\cite{agarwal2022quantifying,demkowicz2015quantum}:

\begin{align}\label{Cramer-Rao Bound}
    \text{Var}(\hat{\omega}_0) \geq \frac{1}{M F(\omega_0)},
\end{align}

 where $F(\omega_0)$ is the classical Fisher information,

\begin{align}\label{Measured Fisher Information}
    F(\omega_0) = \sum_{j} \frac{\left(\partial_{\omega_0} P_{j}\right)^2}{P_j},
\end{align}

and $P_j$ is the probability of obtaining the outcome $j$. The quantum Fisher information for pure probe states encoding the parameter $\omega_0$ through the evolution under control field as $\ket{\psi} = U\ket{0} = a\ket{0} + b\ket{1}$, is expressed as

\begin{align}\label{Predicted Fisher Information}
    F_{\mathcal{Q}}(\omega_0) = 4\left(\bra{\partial_{\omega_0}\psi}\ket{\partial_{\omega_0}\psi} - \abs{\bra{\partial_{\omega_0}\psi}\ket{\psi}}^2\right).
\end{align}

Evaluation of the Fisher information can be difficult and sensitive to approximations so we will evaluate these quantities symbolically using Wolfram Mathematica. The probability of measuring the system in the ground and excited states can be obtained from Eq.~\eqref{Cayley-Klein Parameters} as 

\begin{subequations}\label{Probabilities}
\begin{eqnarray}
 &P_0= \abs{a}^2 = 1 -\sech^2\left(\frac{\pi \tau \Delta}{2}\right)\sin^2\left(\frac{\pi\tau\Omega_0 }{2}\right),\label{Ground State Probability}\\
&P_1 = \abs{b}^2 = \sech^2\left(\frac{\pi \tau \Delta}{2}\right)\sin^2\left(\frac{\pi\tau \Omega_0}{2}\right).
\label{Excited State Probability}
\end{eqnarray}
\end{subequations}

Using these expressions, the classical Fisher information for the parameter $\omega_0$ can be evaluated to be

\begin{align}\label{Single Pulse Classical Fisher Information}
    F(\omega_0) = \frac{2\pi^2\tau^2\sin^2\left(\frac{\pi\tau\Omega_0}{2}\right)\tanh^2\left(\frac{\pi\tau\Delta}{2}\right)}{\cos(\pi\tau\Omega_0)+\cosh(\pi\tau\Delta)},
\end{align}

and for the $\pi$-pulses, $\Omega_0 = 1/\tau$, the variance in $\omega_0$ is lower bounded by

\begin{align}
    \text{Var}(\hat{\omega}_0) \geq \frac{\cosh^2(\frac{\pi\tau\Delta}{2})}{\pi^2 \tau^2M}.
\end{align}

This inequality indicates that the variance decreases with pulse duration consistent with the Fourier narrowing of the spectral linewidth and with the number of measurements, consistent with statistical sampling. 

\begin{figure}
    \centering
    \includegraphics[width=\linewidth]{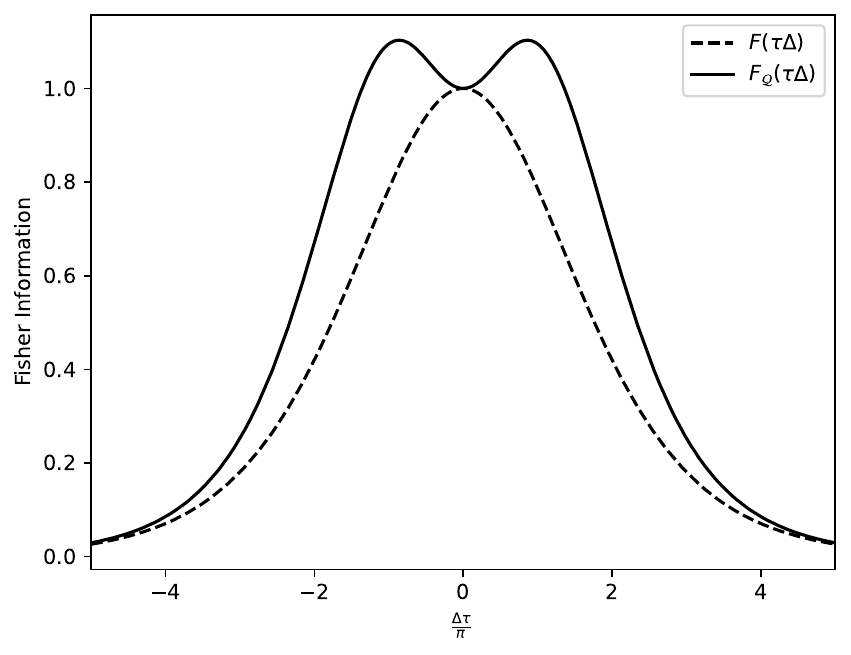}
    \caption{The classical Fisher information $F(\omega_0)$ (dashed) and quantum Fisher information $F_{\mathcal{Q}}(\omega_0)$ (solid) for a single pulse resonance measurement with Rabi frequency $\Omega_0 = 1/\tau$.The quantum Fisher information is greater than the classical Fisher information away from resonance.}
    \label{fig: Single Pulse Fisher Information}
\end{figure}

Figure \ref{fig: Single Pulse Fisher Information} presents a comparison of the quantum Fisher information evaluated from the probability amplitudes in Eq. \eqref{Cayley-Klein Parameters} and the classical Fisher information in Eq. \eqref{Measured Fisher Information}. The two quantities coincide on resonance but the quantum Fisher information is always greater than the classical Fisher information for any value of the detuning away from resonance, meaning the measurement locally saturates the bound on resonance.

Experimentally, the full width at half the maximum (FWHM) of the probability distribution about resonance is a common proxy for variance of the resonance measurement. To assess its performance as an estimator of precision, we will compare the FWHM to the lower bound set by the Fisher information. For the probability distribution in Eq.~\eqref{Probabilities} the FWHM is  

\begin{align}\label{Full Width at Half Maximum}
  \text{arc}P_0\left(\frac{1}{2}\right)_+ - \text{arc}P_0\left(\frac{1}{2}\right)_- = \frac{4\text{arcsinh}(1)}{\pi \tau}, 
\end{align}

where $\text{arc}P_0(\cdot)_{\pm}$ is the inverse probability function returning the value of $\omega$ on the right and left side of the maximum, respectively. The resulting width is consistent with the Cramér-Rao bound in Eq. \eqref{Single Pulse Classical Fisher Information}.

\section{Composite Pulse Resonance\label{Section III}}

\begin{figure}
    \centering
    \includegraphics[width=\linewidth]{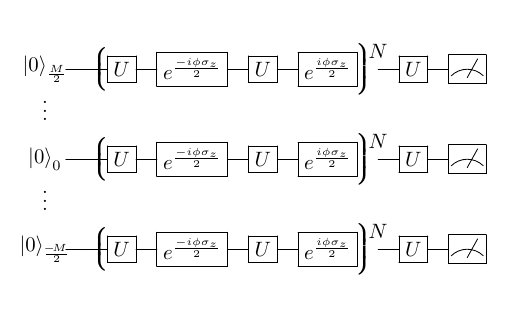}
    \caption{Quantum circuit diagram of the composite pulse resonance measurement. A single experiment consists of a probe state prepared in the ground state interacting with $(2N+1)$ pulses with an alternating phase shift $\phi$ followed by a measurement of the population in the ground state. The index of the final state closest to the ground state yields an estimate of the atomic resonance frequency.}\label{fig:Composite Pulse Sensing Circuit}
\end{figure}

Composite pulse theory was developed in the field of nuclear magnetic resonance to compensate for control field errors by applying sequences of pulses with specific relative phases between pulses \cite{freeman1998spin}. We employ these pulse sequences to make the response of the probe system more sensitive to parameter variations; improving the precision of the measurement. The sequential interaction of N electromagnetic pulses with the probe is described by the $N$th power of the propagator matrix for a single pulse, denoted $U^N$ \cite{vitanov1995coherent}. By Sylvester's theorem, the elements of $U^N$ can be expressed as polynomials in the elements of $U$ \cite{brandi2020composition}, yielding the composite pulse propagator 

\begin{align}
    U^N = \begin{pmatrix}
        T_N +ia_{\mathfrak{I}} U_{N-1}& -b^*U_{N-1}\\
       b U_{N-1} & T_N - ia_{\mathfrak{I}}U_{N-1}
    \end{pmatrix},
\end{align}

where $T_N = T_N(a_{\mathfrak{R}}),\ U_N = U_{N}(a_{\mathfrak{R}})$ are Chebyshev polynomials of the first and second kinds in $\theta = \arccos(a_{\mathfrak{R}})$, with $a_{\mathfrak{R}} = \mathfrak{Re}(a)$ and $a_{\mathfrak{I}} = \mathfrak{Im}(a)$~\cite{mason2002chebyshev}. We will consider a composite pulse sequence composed of $(2N+1)$ hyperbolic secant pulses with a phase shift $\phi$ between subsequent pulses. The total interaction time of the composite pulse will be equal to the single pulse length $\tau$ so the individual pulse length within the sequence is rescaled to $\tau_{2N+1} = \frac{\tau}{(2N+1)}$. This rescales the Rabi frequency for each $\pi$-pulse to $\Omega_{2N+1} = 1/\tau_{2N+1} = (2N+1)/\tau$. The total evolution is then described by the propagator matrix,

\begin{align}
U^{2N+1}\left(\phi\right) = U\left(e^{\frac{i\phi \sigma_z}{2}}Ue^{-\frac{i\phi \sigma_z}{2}}U\right)^N.
\end{align}

This composite propagator will generally depend on the phase shift $\phi$, but we will only investigate the cases $\phi = 0$ and $\phi = \pi$ that are commonly used for rotary spin echos~\cite{solomon1959rotary,vitanov2021quantum}. Physically, such a phase shift can be performed by negating the phase of the applied field on a time scale much shorter than the pulse length.

\begin{figure}
    \centering
    \includegraphics[width=\linewidth]{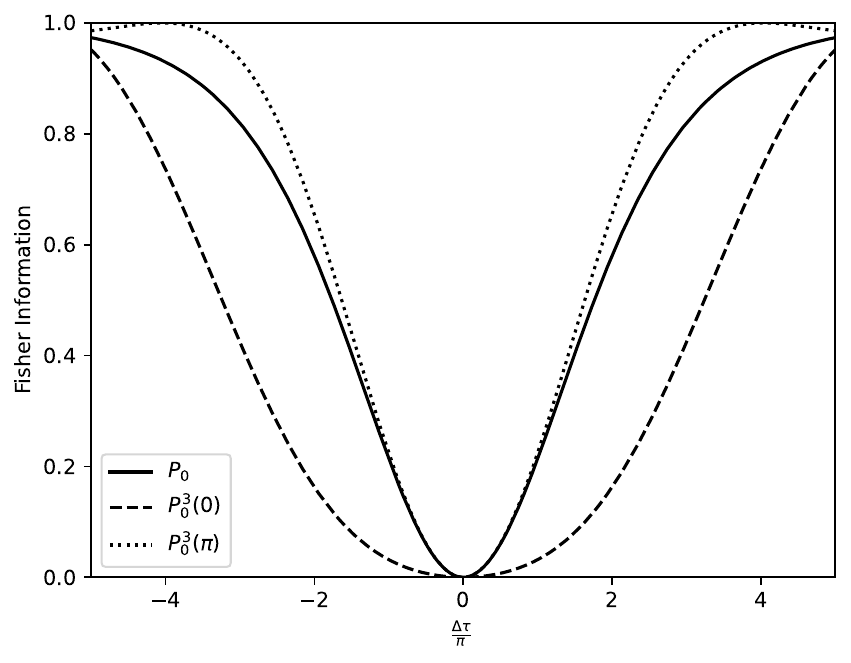}
    \caption{The probability of measuring the population in the ground state for one pulse: $P_0$ (solid) and three pulses: $P_0^{3}(0)$ (dashed), $P_0^{3}(\pi)$ (dotted). All probabilities achieve a minimum at $\omega = \omega_0$ but, the variance is minimized for $\phi = \pi$. }\label{fig: Composite Pulse Probabilities Odd}
\end{figure}

The composite pulse resonance measurement depicted in Figure~\ref{fig:Composite Pulse Sensing Circuit} is conducted in the same manner as the single pulse resonance measurement in Figure~\ref{fig:Single Pulse Sensing Circuit} but with a composite 2N+1 pulse with total duration $\tau$. Figure \ref{fig: Composite Pulse Probabilities Odd} plots, for comparison, the probability of measuring the probe in the ground state for one pulse $P_0$ and three pulses with $\phi = \pi$ phase $P_0^3(\pi)$ and $\phi = 0$ phase $P_0^{3}(0)$. One can observe a significant difference in the variance of the measured response between the single and composite pulse resonance measurements as well as values of the phase. 

Let us proceed by denoting the classical and quantum Fisher information for the atomic resonance frequency $\omega_0$, conditioned on the value of $\phi$ as $F^{2N+1}(\omega_0;\phi)$ and $F_{\mathcal{Q}}^{2N+1}(\omega_0;\phi)$, respectively.
The propagator matrix for the $\phi = \pi$ composite pulse sequence is 

\begin{align}
    U^{2N+1}(\pi) &= \begin{pmatrix}
        a & -b^*\\
        b & a^*
    \end{pmatrix}\begin{pmatrix}
        a^2 + bb^{*} & -2ia_{\mathfrak{I}}b^*\\
        -2ia_{\mathfrak{I}}b & (a^*)^2 + bb^*
    \end{pmatrix}^N\\
     &= \begin{pmatrix}
        a_{\mathfrak{R}}V_N +ia_{\mathfrak{I}} W_{N}& -b^*V_{N}\\
       bV_{N} & a_{\mathfrak{R}}V_N -ia_{\mathfrak{I}} W_{N}
    \end{pmatrix},
\end{align}

where $V_N = V_N(1-a_{\mathfrak{I}}^2)$, $W_N = W_N(1-a_{\mathfrak{I}}^2)$ are Chebyshev polynomials of the third and fourth kind in $\theta_{\pi} = \arccos(1-2a_{\mathfrak{I}}^2)$~\cite{mason2002chebyshev}. The probability of measuring the probe in the ground state is 

\begin{align}\label{Composite Pulse Pi Probability}
    P_0^{2N+1}(\pi) = \abs{b}^2\frac{\cos^2\left(\left(N+\frac{1}{2}\right)\arccos(1-2a_{\mathfrak{I}}^2)\right)}{\cos^2\left(\frac{1}{2}\arccos(1-2a_{\mathfrak{I}}^2)\right)}.
\end{align}

\begin{figure}
    \centering
\includegraphics[width=\linewidth]{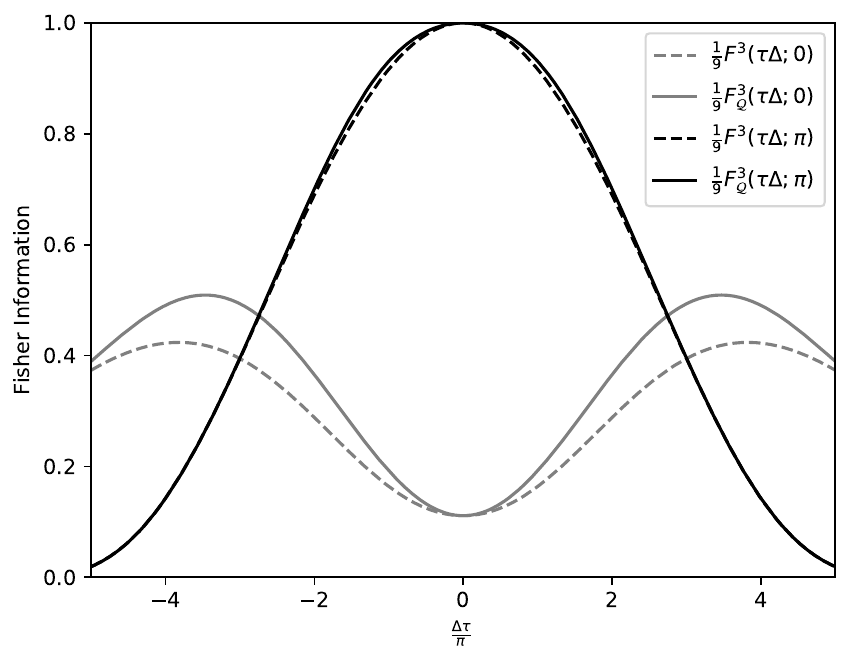}
    \caption{The classical  (dashed) and quantum (solid) Fisher information for $\phi = 0$ (black) and $\phi = \pi$ (grey) with $\tau_{3} = \tau/3$ and $\Omega_3 = 3/\tau$. The classical and quantum Fisher information coincide for all values of the detuning for the composite pulse measurement with $\phi = \pi$.}
    \label{fig:Composite Pulse Fisher Information}
\end{figure}

The classical Fisher information evaluated on resonance is

\begin{align}\label{Pi Fisher Information}
    F^{2N+1}(\omega_0;\pi) = \pi^2\tau^2;
\end{align}

 identical to the single pulse Fisher information in Eq. \eqref{Single Pulse Classical Fisher Information}. However, as one can observe in Figure \ref{fig:Composite Pulse Fisher Information}, the classical and quantum Fisher information are equal for all values of the detuning; indicating that the composite pulse scheme globally saturates the Cram\'er-Rao bound. Thus, the composite pulse method makes optimal use of the full probability distribution for the estimation of the atomic resonance frequency. Following on, the FWHM is exactly $1/\tau$, lower bounded by the Fisher information and independent of $N$.

Now consider $2N+1$ pulses with phase $\phi = 0$ and propagator matrix 

\begin{align}
    U^{2N+1}(0) &= \begin{pmatrix}
        a & -b^*\\
        b & a^*
    \end{pmatrix}\begin{pmatrix}
        a^2 - bb^* & -2a_{\mathfrak{R}}b^*\\
        2a_{\mathfrak{R}}b & (a^*)^2 - bb^*
    \end{pmatrix}^N\\    
     &= \begin{pmatrix}
        a_{\mathfrak{R}}V_{N} +ia_{\mathfrak{I}}W_{N}& -b^*W_{N}\\
       b W_N & a_{\mathfrak{R}}V_{N} -ia_{\mathfrak{I}}W_{N}
    \end{pmatrix}.
\end{align}

The matrix elements now become $V_N = V_N(2a_{\mathcal{R}}^2-1)$ and $W_N = W_N(2a_{\mathcal{R}}^2-1) $; Chebyshev polynomials of the third and fourth kind in $\theta_0 = \arccos(2a_{\mathfrak{R}}^2 - 1)$. The probability of measuring the probe in the ground state is

\begin{align}
    P_0^{2N+1}(0) = \abs{b}^2\frac{\sin^2\left(\left(N+\frac{1}{2}\right)\arccos(2a_{\mathfrak{R}}^2-1)\right)}{\sin^2\left(\frac{1}{2}\arccos(2a_{\mathfrak{R}}^2-1)\right)}.
\end{align}

The resulting classical Fisher information evaluated on resonance is

\begin{align}\label{zero Fisher Information}
    F^{2N+1}(\omega_0;0) = \frac{\pi^2 \tau^2}{(2N+1)^2}.
\end{align}

The Fisher information, and hence also the variance decreases quadratically with increasing $N$. As shown in Figure \ref{fig:Composite Pulse Fisher Information} the quantum Fisher information upper bounds the classical Fisher information everywhere away from resonance. However, the FWHM about resonance is $\sqrt{2N+1}/\sqrt{\log(2)}\tau$, suggesting that the variance increases linearly with $N$. The FWHM does not scale with the Fisher information, overestimating the precision of the measurement. A separate proxy for the difference in precision between the two values of $\phi$ is the second derivative of the probability distribution on resonance. For the case $\phi = \pi$ we find $\partial_{\omega}^2P^{2N+1}(\pi)\eval_{\omega= \omega_0} = (\pi\tau)^2/2$ and for the case $\phi = 0$ we have $\partial_{\omega}^2P^{2N+1}(0)\eval_{\omega= \omega_0} = (\pi\tau)^2/2(2N+1)^2$. Thus, the curvature of the distributions about resonance scales with the Fisher information while the FWHM does not. Identifying an estimator that attains the Cram\'er-Rao bound is a classic problem in estimation theory.

\section{Conclusions}

We have analyzed the limits of precision when estimating the atomic resonance frequency using the simple resonance measurement via the Fisher information. Although the hyperbolic secant–shaped pulse offers maximal signal contrast, due to complete population inversion, the Fisher information reveals the potential for further enhancement in estimation precision. We demonstrate that a simple composite pulse extension of the resonance experiment globally saturates the Cram\'er–Rao bound for all values of the detuning parameter. This result identifies a practical scheme for achieving the fundamental precision bounds attainable with the standard resonance method, without requiring entangled resources or collective measurements. 
It is interesting to note that for large $N$, the largest contribution to the response profile comes from the Chebyshev polynomials of the third and fourth kind. This is a special case of the quantum signal processing theorem and may be generalized to further engineer the system's response for high precise measurements \cite{PRXQuantum.2.040203}. 
During the preparation of this manuscript, we became aware of the preprint by Olivares et al. \cite{olivares2025ultimate}, which investigates the Fisher information for protocols employed in the operation of atomic clocks.

\begin{acknowledgments}
    T.H. would like to thank Akira Sone, Maxim Olchanyi, Victor Bastidas, and Adnan Al-Abar for helpful discussions. T.H. and G.S.A. acknowledge the support of Air Force Office for Scientific Research (Award No. FA-9550-20-10366), the Robert A. Welch Foundation (A-1943-20240404), DOE Award No. DE-AC36-08GO28308 and NSF Award No. 2426699. J.A. was supported in part by the University of Massachusetts Boston's College of Science and Mathematics Dean’s Doctoral Research Fellowship through support from Oracle, project ID R2000000002572.
\end{acknowledgments}

\bibliographystyle{apsrev4-2}
\bibliography{main.bib}

\end{document}